%% file: arxiv.tex
\documentclass[10pt,twocolumn,letterpaper]{article}
\usepackage{cvpr}
\usepackage[accsupp]{axessibility}

\input{sec/preamble}
\definecolor{cvprblue}{rgb}{0.21,0.49,0.74}
\usepackage[pagebackref,breaklinks,colorlinks,allcolors=cvprblue]{hyperref}

\def\paperID{1333}
\def\confName{CVPR}
\def\confYear{2025}

\newcommand{\supplRefMeshEmbeddedGaussians}{\cref{sec:mesh_embedded_gaussians}}
\newcommand{\supplRefTeaser}{\cref{fig:1_teaser}}
\newcommand{\supplRefPbr}{\cref{fig:4_pbr}}
\newcommand{\supplRefExperiments}{\cref{sec:experiments}}
\newcommand{\supplRefBaselines}{\cref{fig:4_baselines}}
\newcommand{\supplRefPbrAppearance}{\cref{sec:pbr_appearance}}

\begin{document}
\title{PGC: Physics-Based Gaussian Cloth from a Single Pose}
\author{
Michelle Guo$^{*1}$ \hspace{0.4em}
Matt Jen-Yuan Chiang$^{2}$ \hspace{0.4em}
Igor Santesteban$^{2}$ \hspace{0.4em}
Nikolaos Sarafianos$^{2}$ \hspace{0.4em} \\
Hsiao-yu Chen$^{2}$ \hspace{0.4em} 
Oshri Halimi$^{2}$ \hspace{0.4em}
Alja\v{z} Bo\v{z}i\v{c}$^{2}$ \hspace{0.4em}
Shunsuke Saito$^{2}$ \hspace{0.4em} 
Jiajun Wu$^{1}$ \hspace{0.4em}
C. Karen Liu$^1$ \hspace{0.4em} \\
Tuur Stuyck$^2$ \hspace{0.4em}
Egor Larionov$^{2}$ \\
{\normalsize $^1$Stanford University}  \quad
{\normalsize $^2$Meta Reality Labs}
}
\input{figtext/1_teaser}

\input{sec/0_abstract}    
\input{sec/1_intro}
\input{sec/2_related}
\input{sec/3_method}

\input{sec/4_experiments}

\input{sec/5_discussion}
\input{sec/6_conclusion}

\clearpage

\section*{Acknowledgements}
We thank S\'ebastien Speierer for providing advice on inverse rendering and Tomas Simon for help with processing captures. This research was supported by Meta Reality Labs. Michelle Guo is also supported by the NSF GRFP, NSF RI \#2211258 and \#2338203, and the Stanford Institute for Human Centered AI (HAI).

\balance
{
    \small
    \bibliographystyle{ieeenat_fullname}
    \bibliography{References}
}

\appendix
\input{sec/X_suppl}

\end{document}

%% file: sec/preamble.tex
\definecolor{MyDarkBlue}{rgb}{0,0.08,1}
\definecolor{MyLightBlue}{RGB}{55, 140, 231}
\definecolor{MyDarkGreen}{rgb}{0.02,0.6,0.02}
\definecolor{MyDarkRed}{rgb}{0.8,0.02,0.02}
\definecolor{MyDarkOrange}{rgb}{0.40,0.2,0.02}
\definecolor{MyLightOrange}{rgb}{0.90,0.5,0.02}
\definecolor{MyPurple}{RGB}{111,0,255}
\definecolor{MyLightPurple}{RGB}{126, 96, 191}
\definecolor{MyRed}{rgb}{1.0,0.0,0.0}
\definecolor{MyGold}{rgb}{0.75,0.6,0.12}
\definecolor{MyDarkgray}{rgb}{0.66, 0.66, 0.66}
\definecolor{MyLightGray}{rgb}{0.8, 0.8, 0.8}
\definecolor{MyWineRed}{rgb}{0.694,0.071, 0.149}
\definecolor{nicegreen}{rgb}{0.1, 0.6, 0.2}

\usepackage{balance}
\usepackage{booktabs}
\usepackage{amsmath}
\usepackage{bm}
\usepackage{lipsum}
\usepackage{multirow}
\usepackage{tabularx}

\newcommand{\myparagraph}[1]{\noindent\textbf{#1}:}

\newcommand{\image}{\bm{I}}
\newcommand{\numcams}{C}
\newcommand{\gtimage}{\image}

\newcommand{\lowpass}{l}
\newcommand{\highpass}{h}

\newcommand{\reconstruction}{\hat{\image}}
\newcommand{\predimage}{\bm{G}}
\newcommand{\gshaded}{\bm{S}}
\newcommand{\alphaimage}{\bm{\alpha}}
\newcommand{\fgmask}{\bm{M}}

\newcommand{\vertices}{\bm{V}}
\newcommand{\faces}{\bm{F}}

\newcommand{\texture}{\bm{T}}
\newcommand{\light}{\bm{L}}

\newcommand{\rasterizer}{\mathcal{G}}
\newcommand{\cg}{\mathcal{R}}

\newcommand{\gopacity}{\alpha}
\newcommand{\gscale}{\bm{s}}
\newcommand{\grotation}{\bm{r}}
\newcommand{\gposition}{\bm{\mu}}
\newcommand{\gsphs}{\phi}
\newcommand{\grgb}{\bm{c}}

\newcommand{\tposition}{\bm{\tau}}
\newcommand{\trotation}{\bm{R}}

\newcommand{\R}{\mathbb{R}}
\newcommand{\Hamilton}{\mathbb{H}}

\newcommand{\myeqref}[1]{\cref{#1}}

\newcommand\blfootnote[1]{%
  \begingroup
  \renewcommand\thefootnote{}\footnote{#1}%
  \addtocounter{footnote}{-1}%
  \endgroup
}

%% file: figtext/1_teaser.tex
\twocolumn[{%
            \renewcommand\twocolumn[1][]{#1}%
            \maketitle
            \centering
            \vspace{-2em}
            \includegraphics[width=0.95\linewidth]{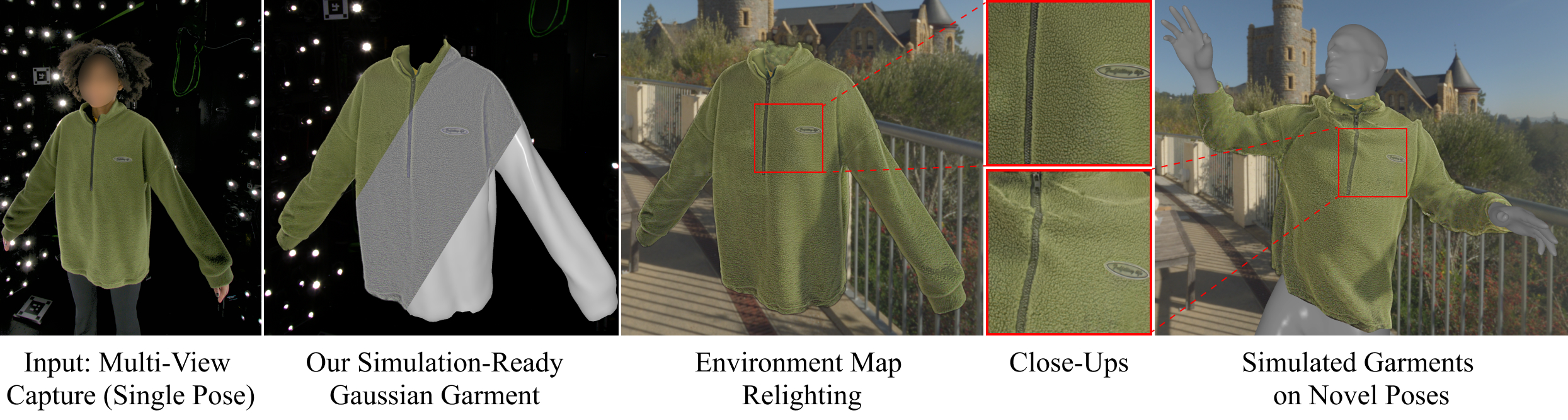}
            \captionof{figure}{
                We present a method to recover photorealistic, simulation-ready garments from a multi-view capture of a \emph{single} static pose. The recovered garments consist of simulatable geometry and fine-detail appearance. Moreover, our results generalize to novel motion as they can be simulated on human motion sequences, and our garments are relightable.
                \vspace{1em}
            }
            \label{fig:1_teaser}
        }]

%% file: sec/0_abstract.tex
\begin{abstract}
We introduce a novel approach to reconstruct simulation-ready garments with intricate appearance. 
Despite recent advancements, existing methods often struggle to balance the need for accurate garment reconstruction with the ability to generalize to new poses and body shapes or require large amounts of data to achieve this.
In contrast, our method only requires a multi-view capture of a single static frame. We represent garments as hybrid mesh-embedded 3D Gaussian splats, where the Gaussians capture near-field shading and high-frequency details, while the mesh encodes far-field albedo and optimized reflectance parameters.  
We achieve novel pose generalization by exploiting the mesh from our hybrid approach, enabling physics-based simulation and surface rendering techniques, while also capturing fine details with Gaussians that accurately reconstruct garment details. 
Our optimized garments can be used for simulating garments on novel poses, and garment relighting. Project page: \href{https://phys-gaussian-cloth.github.io}{phys-gaussian-cloth.github.io}. \blfootnote{*This work was conducted during an internship at Meta Reality Labs}
\end{abstract}

%% file: sec/1_intro.tex
\section{Introduction}
Clothing plays a vital role in how we present ourselves to the world, and it is essential to accurately recreate and represent a person's attire when creating virtual avatars. 
The automatic reconstruction of animatable garments is a vital research area, and it's no wonder that the field has witnessed a surge in interest in garment reconstruction techniques. Over the years, many advances have been made to automatically recreate them from scans~\cite{li2023diffavatar, PhysAavatar24, kim2024gala}, text-guidance~\cite{he2024dresscode, srivastava2025wordrobe}, monocular videos~\cite{qiu2023rec, li2021deep}, RGB-D sensors~\cite{yu2019simulcap}, or even images~\cite{Yang2018GarmentRecovery, sarafianos2024garment3dgen, li2024garmentdreamer, li2024garment}.

Traditionally, garment reconstruction methods have focused on mesh-based representations~\cite{li2023diffavatar, bang2021estimating, xiang2022dressing} due to their natural fit for simulation purposes~\cite{stuyck2022cloth}. Physics-based simulation models allow one to generate natural and physically plausible motions under a variety of body movements. 
However, mesh-based representations are limited to the mesh resolution and often struggle to capture geometric details such as belts, pockets, and zippers or fabrics with significant thickness and fuzziness such as knits or furs. 
On the other hand, recent development of 3D Gaussian splat (3DGS) reconstruction~\cite{kerbl2023Dgs} applied to clothed avatars~\cite{li2024animatablegaussians, lin2024layga, rong2024gaussiangarments} have shown that they are capable of capturing such detailed features resulting in an appearance that is true to the real clothing. Despite significant progress in point-based simulation models for point-based representations of volumetric objects~\cite{zhang2025physdreamer, jiang2024vr, xie2024physgaussian, chen2022virtual}, point-based cloth simulation is considered computationally expensive \cite{yuan2009meshless} but largely under-explored. In addition, existing works tend to require multi-frame tracking~\cite{PhysAavatar24, rong2024gaussiangarments}.
A concurrent work called Gaussian Garments~\cite{rong2024gaussiangarments} also proposes a hybrid representation. Gaussian Garments achieves this by leveraging video sequences of multi-view captures. They use StyleUNet to predict color shifts and lighting effects. In order to train this module, they require tracked sequences of garments in motion. However, tracking highly deformable and dynamic cloth surfaces is computationally expensive and remains a challenging problem. 
Inaccuracy can lead to blurring of the appearance in the final result.
In contrast, we propose a novel appearance model that reconstructs the appearance from a single time instance only, eliminating the need for temporally tracking deformable surfaces, leading to a less cumbersome and more efficient method.

The goal of this work is to reconstruct the appearance of a garment from a multi-view capture of a \textit{single frame}. A natural approach to this problem is to leverage the recent work on 3D Gaussian splat (3DGS) scene reconstruction \cite{kerbl2023Dgs}, which uses a set of 3D Gaussian splats rendered from multiple views to fit a set of captures. Each 3D Gaussian splat is essentially a point with a local coordinate frame and associated view-dependent appearance properties rendered with a soft falloff modelled using a Gaussian kernel.
With this approach we capture the fine volumetric details on the surface of fabrics like pilling and fly-aways that give many materials their characteristic look.

A limitation of the original approach~\cite{kerbl2023Dgs} is that splats are not constrained to specific locations in space, which can lead to severe visually jarring artifacts when their positions or rotations are perturbed. To alleviate this issue, recent works employed a mesh-embedded
splat representation~\cite{guedon2023sugar, qian2024gaussianavatars, rong2024gaussiangarments, moon2024expressive} that anchors splats on a triangle mesh to better control their position and scale, which allows them to follow the deforming mesh more faithfully. We adopt this approach, recognizing the advantage of this representation as it allows for physics-based deformations of the underlying mesh using a simulation pipeline.

Another difficulty with Gaussian splatting is that splats overfit to the scene appearance, baking in shading effects like lighting conditions, specular highlights, and self-shadowing. This limits their use beyond mere reconstruction.  
We propose an approach to address this limitation in garment assets generated from multi-view single pose garment captures. 
Instead of relying solely on Gaussian splat optimization to reconstruct the static scene, we exploit the strengths of traditional physically-based rendering (PBR) techniques to reproduce far-field pose-dependent shading, and use 3DGS reconstruction to fill in the missing pose-independent details such that the final result remains faithful in appearance to the original garment. This decomposition of responsibilities based on representation fits well with frequency-based decomposition of signals.

To generate garments in novel poses, we employ a physics-based cloth simulator to generate a deformed garment mesh. Using the deformed mesh, our method combines the high frequency information of the reconstructed mesh-embedded 3DGS with the low frequency information obtained from the mesh-based representation, which is obtained through physics-based rendering of the mesh with its reconstructed texture map and reflectance parameters. 

To summarize, we propose the following contributions:
\begin{itemize}
    \item The first method to reconstruct high quality simulation-ready garments leveraging only a \textit{single frame} of a multi-view capture.
    \item A hybrid garment rendering method, which leverages both the PBR-based shading for far-field effects and Gaussian splats for reproducing near-field fine details on the surface of fabrics.
    \item We evaluate the sheen's impact on far-field garment appearance and achieve better reconstruction results with a cloth-specific PBR model.
\end{itemize}

%% file: sec/2_related.tex
\vspace{-0.1cm}
\section{Related Work}
\vspace{-0.1cm}
\noindent\textbf{Mesh-based Garment Reconstruction}:  
Garment reconstruction methods have tackled this problem using specialized patterns to track cloth geometry accurately~\cite{halimi2022pattern, chen2021capturing, white2007capturing, scholz2005garment}, but the reconstruction is limited to garments manufactured with a specific pattern thus limiting the ability to capture appearance. Recent work~\cite{zhan2024pattern} demonstrated the possibility of synthetically augmenting appearance for captured patterned clothing.  
A plethora of mesh-based works have been introduced recently that approach the problem from a deformation~\cite{sarafianos2024garment3dgen, xiang2022dressing},  differentiable cloth simulation~\cite{stuyck2023diffxpbd, li2022diffcloth, li2023diffavatar, grigorev2023hood}, and inverse rendering ~\cite{PhysAavatar24, wang24intrinsicavatar} perspectives. 
However, mesh-based works are unable to represent high-frequency details on the surface of the garment that are challenging to model with a mesh (\eg fur, knits, pockets). In contrast, we augment our mesh garments with Gaussians to model the appearance of near-field geometry.

\input{figtext/3_method}

\noindent\textbf{Clothed Gaussian Avatars}: Several methods~\cite{lin2024layga, qian20233dgsavatar, peng2024pica, rong2024gaussiangarments, wang24intrinsicavatar} have been proposed to animate cloth avatars using 3D Gaussian splats, including learning deformations based on the underlying avatar movements. D3GA~\cite{Zielonka2023Drivable3D} embeds 3D Gaussians onto a tetrahedral cage driven by joints and keypoints, while Animatable Gaussians (AG)~\cite{li2024animatablegaussians} predict pose-dependent Gaussian offsets on 2D position maps and use linear blend skinning to animate the Gaussians. 
However, these approaches require multi-frame input and often struggle with tracking garments due to the lack of distinctive texture cues and self-occlusion in clothing. This can lead to a degradation in appearance. In contrast, our method overcomes these limitations by eliminating the need for tracking and requiring only a single static frame.

\noindent\textbf{Hybrid Approaches}: 
The key motivation behind leveraging hybrid approaches for clothed human applications is that they combine the best of both worlds. Mesh-based approaches are fast and easy to render but fail to capture loose clothing and fine-level details whereas Gaussian splat-based approaches generate hiqh-quality results but their ability to be animated and driven remains an active research problem. 
One line of work~\cite{kirschstein2024gghead, xu2024gaussian, wang2024mega, xu2024gphmv2} has focused on Gaussian heads driven by underlying mesh-based representations. 
In addition, a wide-range of works~\cite{yuan2024gavatar, svitov2024haha, wang2024hisr, pan2024humansplat, pesavento2024anim, feng2022capturing, moon2024expressive} has introduced hybrid approaches for modeling the full clothed body. 
Unlike such approaches that rely on monocular video inputs, we focus on extracting the fine details from a single-pose multi-view setup. Furthermore, our main idea is to leverage Gaussians to fit only the high-frequency details of fabrics, while leaving low-frequency details to the physics-based mesh rendering approach.

\noindent\textbf{Gaussians from Single Pose}:
Several works~\cite{xie2024physgaussian, qureshi2024splatsim, zhang2025physdreamer} have investigated reconstructing objects given a single multi-view observation for a variety of applications. SplatSim~\cite{qureshi2024splatsim} leverages Gaussian splat reconstruction for rigid body objects to improve robot manipulation tasks while 
PhysDreamer~\cite{zhang2025physdreamer} optimizes Gaussians from a multi-view capture of a scene. 
However, they have baked-in appearance and their method is not applicable to garments. Unlike existing works, we reconstruct the Gaussians modeling appearance details as well as the underlying surface mesh from a single pose, enabling realistic garment simulation.

%% file: figtext/3_method.tex
\begin{figure*}[t]
  \centering
  \includegraphics[width=0.96\linewidth]{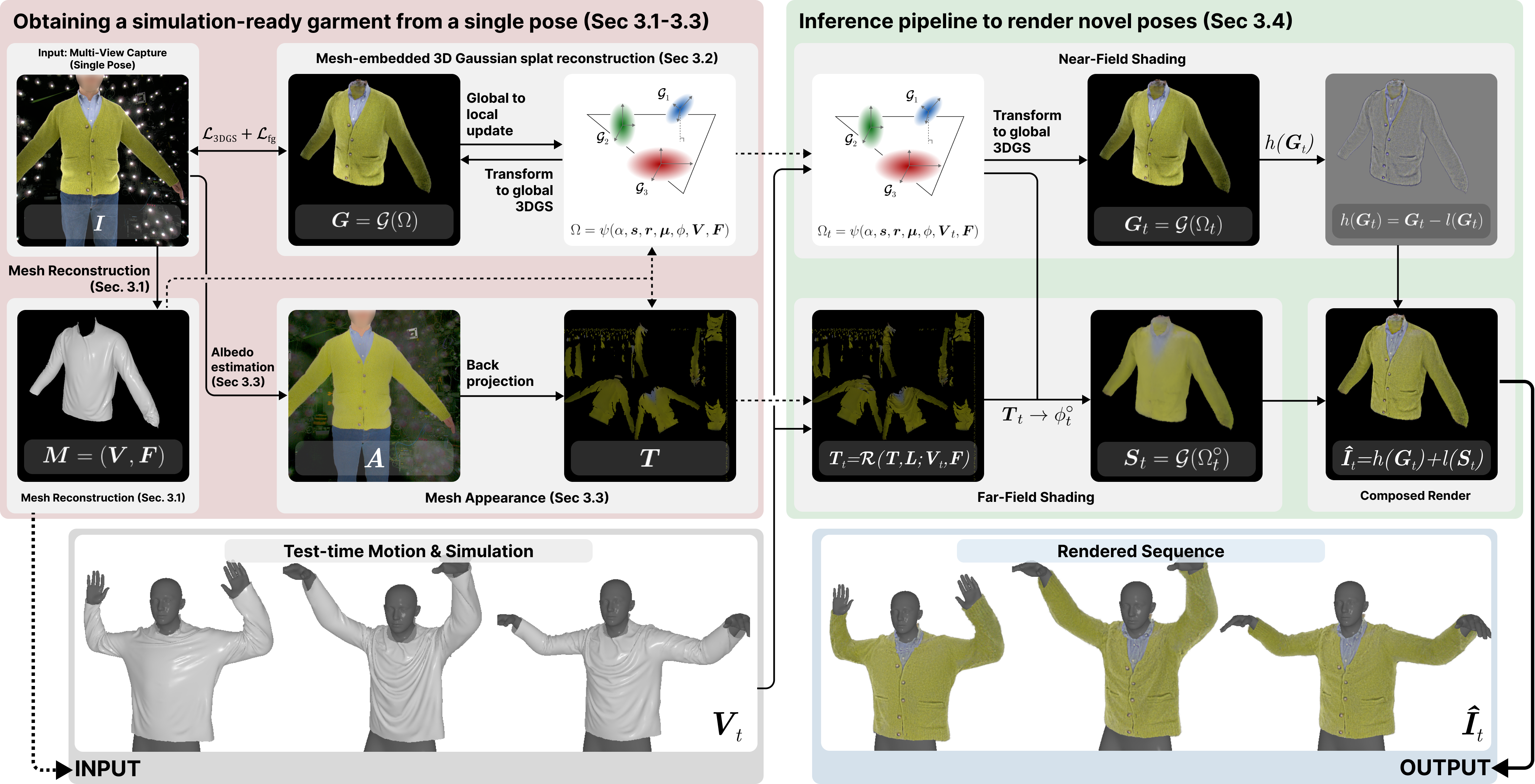}
   \caption{\textbf{Method Overview}. Given a multi-view capture of a clothed human in a single pose, we first extract the garment mesh and fit a 3DGS representation with mesh-embedded Gaussian splats. Simultaneously, we fit an albedo map of the ground truth image and back-project onto the mesh to generate a textured mesh. 
   At inference, the mesh is shaded with a physically-based shading model and the resulting mesh colors are then transferred to zero-order spherical harmonics on the pre-optimized splats. Finally, we combine the high-pass of the original Gaussian splat reconstruction with the low-pass of the traditionally shaded result to produce the final render.}
   \label{fig:3_method}
   \vspace{-0.2cm}
\end{figure*}

%% file: sec/3_method.tex
\vspace{-0.1cm}
\section{Method}
We propose a novel method for reconstructing photorealistic garment appearance from multi-view \textit{single-pose} capture. Our system (\cref{fig:3_method}) consists of three key components: 1. Mesh-embedded 3DGS reconstruction (\cref{sec:mesh_embedded_gaussians}) — building on previous work, we create a mesh-embedded 3DGS representation from our capture. It leverages the strength of 3DGS in reconstructing fine volumetric fabric details while anchoring them on a mesh to enable more faithful deformations; 2. Physics-based rendering (PBR) \cite{pharr2023physically} appearance reconstruction (\cref{sec:pbr_appearance}) — we propose a novel cloth-specific physically-based shading model to reconstruct PBR appearance from real world captures. This accurately captures shadowing and indirect illumination effects under novel pose and lighting conditions; 3. Gaussian-PBR hybrid rendering (\cref{sec:hybrid_rendering}) — we combine the strength of physically-based rendering for far-field pose-dependent shading, and use 3DGS reconstruction to fill in the missing near-field pose-independent details.

\subsection{Preliminaries}\label{ssec:prelim}

We represent garments with a hybrid representation that integrates a mesh with Gaussian splats. 
The splats' positions are defined as offsets from the mesh. This allows the mesh to aid in simulating new dynamics and predicting shading, while the Gaussian splats are used to capture and predict high-frequency volumetric details in the appearance of the simulated surface. In this section, we discuss geometric optimization for mesh-embedded Gaussian splatting and review physically-based rendering used to produce the far-field shading effects.

\noindent\textbf{Mesh Reconstruction}:
Given a set of multi-view images $\{\image_k\}_{k=1}^\numcams$ from $\numcams$ camera views of a clothed human in a static pose, we reconstruct a triangle mesh $(\vertices,\faces)$ with vertex positions $\vertices$ and triangles $\faces$.
We select a near A-pose frame such that the garment's surface is visible, and has minimal wrinkles.
The images are then processed to produce a noisy point-cloud via stereo matching~\cite{bleyer2011patchmatch}.
We further reconstruct the surface~\cite{kazhdan2013screened} using the generated point-cloud, and apply remeshing \cite{sidefx24} to improve the triangle quality making the mesh suitable for simulation.
From the reconstructed mesh of the whole body, we automatically segment out the garment mesh using image-based segmentation~\cite{fu2019imp} and project the maps back to the mesh.

\noindent\textbf{Mesh-embedded Gaussian Splat Representation}:
To couple the Gaussians with the mesh, each Gaussian is mapped to a triangle on the mesh via a triangle-local coordinate frame centered at the centroid of each triangle \cite{qian2024gaussianavatars}.
The orientation of this frame is determined by the triangle's normal, one of the edges normalized, and their cross product.
Each 3D Gaussian is parameterized by its position $\gposition \in \R^3$, rotation quaternion $\grotation\in\Hamilton$ and scale $\gscale\in\R^3_+$ relative to this frame
as well as the coefficients of three degrees of spherical harmonics $\gsphs\in[0,1]^{16\times3}$ and opacity $\gopacity\in\R$. The triangle-local splat positions and rotations are transformed into world space (\cref{sec:mesh_embedded_gaussians}) giving a set of global splat parameters
$\Omega = \psi(\gopacity, \gscale, \grotation, \gposition, \gsphs; \vertices, \faces)$,
before the splats are finally rendered producing the final image
$\predimage = \rasterizer(\Omega)$,
where $\rasterizer$ is the Gaussian rasterizer \cite{kerbl2023Dgs}. The view parameters are implicit in $\rasterizer$.

\subsection{Mesh-Embedded 3DGS Reconstruction}\label{sec:mesh_embedded_gaussians}
Each garment is rendered using a mesh-embedded 3D Gaussian splat representation. In addition to prior work, we further exploit the underlying mesh representation to model far-field lighting and self-shadowing effects using PBR techniques.
The pre-computed asset consists of a 3D Gaussian splat reconstruction of the garment and a corresponding mesh with an albedo texture
estimated from real world captures. These two representations are later combined to render the fine result as described in \cref{sec:hybrid_rendering}.

\noindent\textbf{Gaussian Appearance}:
We sample 1 million Gaussians on the surface of the mesh via triangle point picking
\cite{weisstein24}, which defines the local position $\gposition$ of each Gaussian. We initialize the local rotation $\grotation$, scale $\gscale$ and opacity $\gopacity$ to identity, 1 and 0.1, respectively. Spherical harmonics coefficients $\gsphs$ are initialized randomly. The parameters are optimized through minimizing the sum of two losses. First, the following loss \cite{kerbl2023Dgs} between the reference and rendered images:
\begin{equation}
    \mathcal{L}_{\text{3DGS}} = \lambda \left\| \gtimage_k - \predimage_k \right\|_1 + (1 - \lambda) \text{SSIM}(\gtimage_k, \predimage_k),
\end{equation}
where $\predimage_k$ is the rendered image, $\image_k$ is the reference image for camera $k$ and SSIM is the structural similarity index~\cite{zhou2004image}. 
Following~\cite{kerbl2023Dgs}, we use $\lambda = 0.8$ for all our experiments.
To discourage outliers (\ie Gaussian splats away from the surface that may fit details not related to the fabric), we adapt a regularization loss from NeRF \cite{bi2020nerf} to penalize splats far away from the surface:
\setlength{\abovedisplayskip}{4pt}
\setlength{\belowdisplayskip}{4pt}
\begin{equation}
    \mathcal{L}_{\text{fg}} = \lambda_{\text{fg}}  \left\| \fgmask_k - \alphaimage_k \right\|_1,
\end{equation}
where $\fgmask_k$ is the foreground mask of the ground truth image $\image_k$ obtained using Sapiens~\cite{rawal24sapiens} and $\alphaimage_k$ is the alpha channel of the rendered Gaussians.
This encourages foreground and background Gaussians to have accumulated opacity close to 1 and 0 respectively. 
We use $\lambda_{\text{fg}}=0.1$ in our optimization.
Unlike \cite{rong2024gaussiangarments}, which trains on segmented garment images, we train Gaussians on entire images. To do so, we use $\alphaimage_k$ to blend the rendered garment with images of the empty capture dome.
This allows us to learn fuzzy details on the garment boundaries instead of being limited by the segmentation model accuracy. Note that the alpha regularization loss is also sensitive to the segmentation accuracy and can negatively affect the reconstruction of fuzzy details. 
We alleviate this issue by not enforcing the loss at the contours of the garment, which we automatically identify by eroding and dilating the foreground mask $\fgmask_k$, and masking the pixels at the exclusion. After optimizing Gaussians on the full body, we only keep the Gaussians that are bound to mesh triangles belonging to the segmented garment.

\subsection{PBR Appearance Reconstruction}\label{sec:pbr_appearance}

While the optimized 3D Gaussians (\cref{sec:mesh_embedded_gaussians}) are capable of reconstructing the garment appearance well, especially for the near-field high-frequency fabric details, they contain baked-in lighting from the training frame, and thus novel deformations (\eg, new wrinkles) would not be shaded correctly as demonstrated in~\cref{fig:4_ablations}(b).
Previous works typically rely on video sequences to reproduce Gaussian parameters for different mesh configurations~\cite{rong2024gaussiangarments,PhysAavatar24,peng2024pica} and involve multi-frame tracking. Instead, we leverage physically-based rendering, which naturally accounts for lighting effects such as shadowing and indirect illumination under mesh deformation. This considerably simplifies the pipeline and compute load.
In this section, we detail how we reconstruct a PBR representation from our capture. We first introduce our cloth-specific PBR shading model that is suitable for both cloth appearance modeling and reconstruction. Then, we estimate the garment color using a learning-based approach, removing baked-in lighting from the input images. The other shading parameters of our model are estimated with physically-based differentiable rendering. Finally, together with the reconstructed mesh (\cref{sec:mesh_embedded_gaussians}), such PBR-based representation can be rendered under novel pose, view, and lighting conditions. The relighting capability is shown in \cref{fig:4_relighting}.

\noindent\textbf{Cloth Appearance Model}: The Lambertian model has been commonly used in previous work~\cite{PhysAavatar24}. However, as shown in \cref{fig:4_pbr}(b), this naive view-independent approach misses several key fabric appearance properties and does not fit the actual data well. In contrast, we base our model on the Disney Bidirectional Reflectance Distribution Function~\cite{burley2012physically} (BRDF) $f_{d}$ , a versatile PBR model that is capable of creating realistic appearance for general hard surfaces and beyond. 
Garments exhibit strong  backward and forward scattering at grazing angles, due to their flyaway fibers. This is apparent for fabrics constructed from loose staple yarns such as the fleece and cardigan in \cref{fig:4_pbr}(a). This effect is commonly known as \textit{sheen}. 
We observed that the sheen component of~\cite{burley2012physically} is unable to match the capture and instead adopt the sheen BRDF $f_{s}$ proposed by~\cite{zeltner2022practical}. It is modeled based on optical simulation and takes the important fiber multiple scattering into account. Combining two BRDFs, our final appearance model $f$ results in:
\begin{equation}
\label{eq:brdf}
    f(\bm{x},\bm{i},\bm{o}) = \sigma_{s} f_{s}(\bm{x},\bm{i},\bm{o}) + H_{s}(\bm{o}) \sigma_{d}(\bm{x}) f_{d}(\bm{x},\bm{i},\bm{o}),
\end{equation}
where $\bm{x}$ is the position and $(\bm{i},\bm{o})$ are the lighting and viewing directions. $\sigma_{s}$ is the constant sheen color and $\sigma_{d}(\bm{x})$ is the spatially-varying base color. $H_s(\bm{o})$ is the view-dependent sheen transmission \cite{zeltner2022practical}, which ensures we combine two BRDFs in an energy conserving way. We found that the render using our model matches the input image better compared to both Lambertian model (completely lack of sheen) and Disney BRDF (only having forward scattering component of sheen and lack of roughness control) as demonstrated in \cref{fig:4_pbr}. 

\input{figtext/4_pbr}

\noindent\textbf{Base Color Reconstruction}:
To reconstruct the base color $\sigma_{d}$ of our model for PBR shading, our goal is to extract a spatially-varying lighting-less albedo map from input views of the training frame. 
To remove baked-in lighting, we use a pre-trained intrinsic image decomposition network~\cite{careagaIntrinsic} to predict the albedo for each view. 
Intrinsic image decomposition is inherently under-constrained and a common problem is multi-view inconsistency. 
To address this, we calculate a color correction matrix per view that minimizes the difference between the input image $\gtimage_k$ and the predicted albedo within the segmentation mask $\fgmask_k$. We found that this normalizes the predicted albedo across views effectively. 
We estimate and normalize the albedo from 10 input views, then back-project and merge them in texture space.

\noindent\textbf{Shading Reconstruction}:
Given the estimated base color and the carefully calibrated capture conditions, we use Mitsuba~\cite{Mitsuba3}, a physically-based differentiable renderer, to obtain the other stationary parameters of our model from our training frame. 
The parameters optimized are \textit{roughness}, \textit{sheen color} and \textit{sheen roughness}. To better constrain our model in the cloth appearance space, we set all other parameters as the configuration of a rough dielectric material, including setting \textit{metallic} and \textit{clearcoat} to 0. We optimize the parameters jointly across all input views of the training frame. 
We observed that together with the reconstructed base color, our PBR-based method produces a faithful far-field reconstruction for both training and test frames. 

\subsection{Gaussian---PBR Hybrid Rendering}\label{sec:hybrid_rendering}

We propose a novel approach that combines the strengths of 3DGS (\cref{sec:mesh_embedded_gaussians}) and Physically-Based Rendering (PBR) representations (\cref{sec:pbr_appearance}). 
The former excels at capturing fine volumetric details on fabric surfaces, but struggles with pose-dependent effects, while the latter accurately renders far-field lighting effects with novel poses, but is limited by the underlying mesh quality. 
By decomposing an image into high-frequency and low-frequency components, we approximate the former using Gaussian-based representation and the latter using PBR-based shading. 
Our key observation is that pose-dependent effects like shadowing and indirect illumination are mostly in the low-frequency signals of a render and can be well-approximated by PBR-based shading, while high-frequency signals corresponding to fine fabric details remain largely pose-independent and can be effectively captured by a Gaussian-based representation. 
To be specific, a $W\times H$ RGB image $\image \in [0,1]^{W \times H \times 3}$ can be naturally decomposed into low- and high- frequency components as
\begin{align}
\label{eq:decompose}
    \image = h(\image) + l(\image),
\end{align}
where $\lowpass(\image)$ and $\highpass(\image)$ are a low-pass and high-pass filter on image $\image$ respectively. Our hybrid approach will approximate $h(\image)$ with our 3D-Gaussian representation (we call \textit{near-field shading}) and $l(\image)$ with our PBR model (we call \textit{far-field shading}) and then combine them into a final render. With this, we generalize our garment rendering to novel deformation using single-pose reconstruction. We are able to reproduce both high-frequency fabric details and pose-dependent lighting effects. We demonstrate our results in \cref{sec:experiments}. 
We discuss implementation details below.

\noindent\textbf{Far-field Shading}: 
To create a deformed mesh, we use a physics simulator~\cite{macklin2016xpbd} to generate a set of deformed vertices at time $t$, denoted by $\vertices_t$.
It is natural to render the mesh with a PBR model directly into screen space~\cite{Mitsuba3}. However, mesh rasterization requires a substantially different rendering pipeline than Gaussian splatting. To combine the two representations at run-time, we choose to render into texture space.  To be specific, we render the mesh using our PBR workflow (introduced in \cref{sec:pbr_appearance}) denoted by $\cg$ with reconstructed parameters $\texture$ (including base color and other shading parameters) and a given lighting configuration $\light$. This produces a new shaded texture: \(\texture_t = \cg(\texture, \light; \vertices_t, \faces)\). Our cloth-specific PBR shading model \cref{eq:brdf} and viewing information are implicit in $\cg$. 
For each Gaussian splat, we find the color on $\texture_t$ using the pre-computed correspondence (introduced in \cref{ssec:prelim}) to obtain the zero-order spherical harmonics $\gsphs^{\circ}_t\in\R^3$. Thus, we can use the existing 3DGS renderer $\rasterizer$ to produce a final image using the new colors, but with all other parameters unchanged from the original 3DGS optimization as follows:
\begin{align}
    \gshaded_t &= \rasterizer(\Omega^{\circ}_t), \label{eq:far_field} \\
    \Omega^{\circ}_t &= \psi(\gopacity, \gscale, \grotation, \gposition, \gsphs^{\circ}_t; \vertices_t, \faces). \notag
\end{align}
In contrast to directly rendering Gaussians, this allows us to reconstruct wrinkle details under novel deformations.


\noindent\textbf{Near-field Shading}:
As mentioned earlier, PBR shading lacks many high-frequency details of textured fabric like flyaways.
To recover these details, we first render the full 3DGS reconstruction at the new configuration
\begin{align}
    \predimage_t &= \rasterizer(\Omega_t) \label{eq:near_field}\\
    \Omega_t &= \psi(\gopacity, \gscale, \grotation, \gposition, \gsphs; \vertices_t, \faces). \notag
\end{align}
We then extract the near-field shading and appearance from the optimized 3D Gaussians by applying a high-pass filter $\highpass$ to the Gaussian rendering: \(\highpass(\predimage_t) = \predimage_t - \lowpass(\predimage_t)\),
where $\lowpass(\image)$ is an $\bm{\alpha}$-scaled low-pass filter on image $\image$ where $\bm{\alpha}$ the alpha channel of $\image$.
The low-pass is computed with an alpha-aware Gaussian blur, so that the background does not contaminate the garment color at the boundary.
To achieve this, we renormalize the kernel after scaling it by alpha as $K(x,y) \alphaimage(x,y)$ for each kernel pixel at position $(x,y)$. 
The alpha weighting is applied before kernel normalization. We use the rasterized Gaussian opacity as our alpha mask.

\input{figtext/3_reconstruction}

\noindent\textbf{Composed Render}:
Our final render $\reconstruction$ combines the high-pass near-field reconstruction of \myeqref{eq:near_field} with a low-pass of the corresponding far-field reconstruction from \myeqref{eq:far_field}:
\begin{align}
\reconstruction_t = \highpass(\predimage_t) + \lowpass(\gshaded_t).
\end{align}
\cref{fig:4_decomposition} demonstrates the composed render on 4 different garments.
Together with \myeqref{eq:decompose} we identify the reconstruction error $|\image-\reconstruction_t|$ of our hybrid approach for the training frame as shown in \cref{fig:3_reconstruction} and empirically find the size of the blur kernel that minimizes the reconstruction error. Please refer to the supplemental material for more discussion.

%% file: figtext/4_pbr.tex
\begin{figure}[t]
  \centering
  \includegraphics[width=0.98\linewidth]{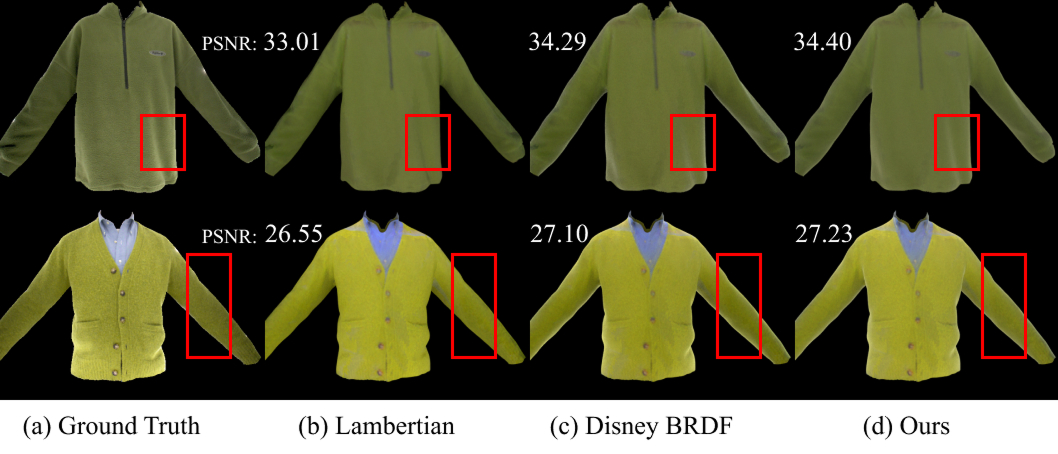}
  \caption{\textbf{PBR model evaluation}.
  Each mesh is shaded with 3 different shading models and compared with ground truth (a). 
  Notice the sheen around the cardigan sleeve and the wrinkle near the fleece abdomen denoted by the red square. 
  The Lambertian model (b) shows no sheen, Disney BRDF (c) over-estimates the forward scattered sheen, whereas our model (d) produces the best match to ground truth sheen. 
  PSNR values are computed to quantitatively demonstrate the improvement of our model over prior methods.}
   \label{fig:4_pbr}
   \vspace{-0.2cm}
\end{figure}

%% file: figtext/3_reconstruction.tex
\begin{figure}[t]
  \centering
  \includegraphics[width=0.98\linewidth]{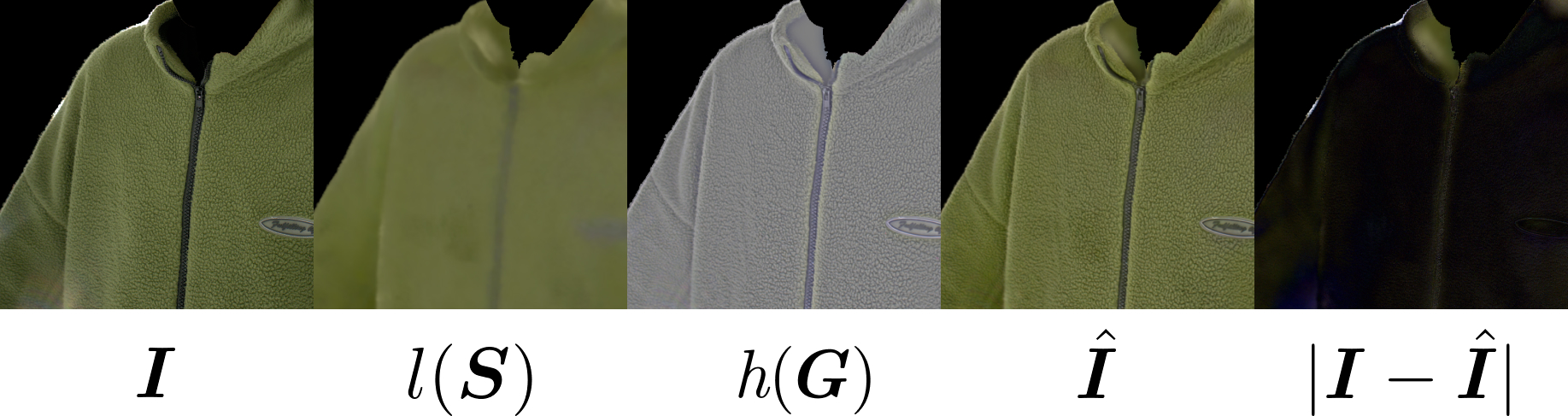}
   \caption{\textbf{Reconstruction.} The fleece garment $\image$ is reconstructed from the sum of the far-field approximation $l(\gshaded)$ and near-field approximation $h(\predimage)$ to produce the final reconstruction $\reconstruction$. The image on the right shows the reconstruction error.}
   \label{fig:3_reconstruction}
\end{figure}

%% file: sec/4_experiments.tex
\vspace{-0.2cm}
\section{Experiments}\label{sec:experiments}

\noindent\textbf{Data}:  
We use four garments: a loose t-shirt, a loose dress, a fleece quarter-zip, and a knit cardigan to demonstrate our ability to handle loose garments and fuzzy fabrics.
The clothed subjects are recorded with a multi-view system consisting of 170 cameras arranged uniformly in a half-dome configuration above the ground. 
Images are captured at $4096\times2668$ resolution.
Each subject followed a predefined script that includes a variety of poses to ensure diverse data capture.
However, despite capturing temporal sequences, we utilize only a single frame near the A-pose for each subject to reconstruct the garments. The rest of the ground truth captures represent novel poses, which we use to evaluate our method in all figures except \cref{fig:4_pbr} and \cref{fig:3_reconstruction}.

\noindent\textbf{Metrics}:
We validate our method using two metrics,
learned perceptual similarity (LPIPS)~\cite{zhang2018unreasonable}
and feature similarity (FSIM)~\cite{zhang2011fsim}, which compares salient low-level features to better match the human visual system. 
Note that pixel-based metrics such as PSNR are less appropriate for our work at test time since we use a simulated result with manually defined parameters that may not align with ground truth as can be seen in \cref{fig:4_baselines} (see \cref{ssec:limitations} for proposed improvements). 
The goal of our approach is not to achieve perfect pixel-to-pixel correspondence at test time. 
Instead, we show that new wrinkles are generated under new deformations and their aesthetic is preserved, which we demonstrate qualitatively in \cref{fig:4_ablations} and \cref{fig:4_baselines}.
Quantitative metrics were computed on four video sequences containing 150 unseen frames each, including a goal-post pose, spooky pose and upside-down U pose as shown in \cref{fig:1_teaser}.
\input{table/4_metrics}
\subsection{Ablation Studies}

\input{figtext/4_ablations}

We compare our method qualitatively and quantitatively with a series of ablations where the importance of key components of our method are evaluated. We provide qualitative comparisons in \cref{fig:4_ablations} as well as quantitative comparisons in \cref{tab:4_metrics} (bottom). Specifically, we explore the performance of: a) 3DGS-Only that has baked in shading from the training pose, which makes the appearance unrealistic when deformed to novel poses, b) PBR-Only that has reshading, but lacks the detailed shading and volumetric flyaways meaning that it can only render 2D surfaces, and c) our approach that combines the far-field relightable BRDF render with the near-field shading and fine volumetric details of the mesh-embedded 3DGS.

\input{figtext/4_decomposition}

\noindent\textbf{Relighting}: We demonstrate that our garments can be rendered under a variety of lighting conditions in \cref{fig:4_relighting}.  Note that compared to pure 3DGS-based methods that require learning pose-dependent intrinsic maps from multi-view multi-pose data, we rely on a single pose only. This figure also demonstrates a clear sheen effect in novel lighting optimized against real world captures.

\subsection{Body Animation \& Cloth Simulation}
Although any simulation method would be effective, we simulate the garment on body mesh sequences using an implementation of the eXtended Position Based Dynamics~(XPBD)~\cite{macklin2016xpbd} simulation method, chosen for its excellent performance characteristics.
We use a SMPL-like parameterized statistical body model.
Given the tracked skeleton motion from the video sequence, we use linear blend skinning (LBS) to animate the human body.
The garment is simulated with the body mesh as the collider.
The garment meshes are directly used for rendering the novel frames.

\subsection{Comparisons with State of the Art}
We compare with two top-performing recent approaches: Animatable Gaussians (AG)~\cite{li2024animatablegaussians} and SCARF~\cite{Feng2022scarf} and provide our qualitative and quantitative comparisons in \cref{fig:4_baselines} and \cref{tab:4_metrics}. We refer the reader to the supplementary material for additional details for each baseline method. 
We observe that both past works struggle on loose clothing like the dress. AG has difficulty modeling loose garments since the underlying skinning module cannot represent dynamic motions, so it produces blurriness at the loose part of the clothing. Since neither baselines use physics simulation to drive the dress motion, the resulting dynamical movements are unrealistic in comparison to ours. 
However, despite producing higher quality motion, our approach does not optimize for garment rest shape and instead uses the reconstructed geometry as the rest shape for simulation. 
This is a well-known phenomenon (\ie sagging~\cite{Hsu2022}), which causes the dress to sag under gravity and hence the patterns on the dress (although more crisp) appear lower in the frame than in the reference. 
We believe this is the main reason for AG to produce slightly higher 
PSNR values and we note that other state-of-the-art methods \cite{rong2024gaussiangarments} excluded quantitative comparisons to AG altogether. Additionally, our shading model also provides better generalization.

Although we cannot explicitly compare results against Gaussian Garments~\cite{rong2024gaussiangarments}, since their code is not available, we note that not unlike their method, we also outperform SCARF in all metrics, without the need for multi-frame tracking. As a result, our method produces more crisp details since we can completely avoid motion-blur effects in video captures and averaging artifacts during multi-frame tracking. In terms of performance, Gaussian Garments requires an additional 24.5 hours for multi-view registration, which our method bypasses entirely.
\input{figtext/4_baselines}

%% file: table/4_metrics.tex
\begin{table}[t]
\centering
\caption{\textbf{Quantitative Comparisons}: Our proposed approach improves significantly upon top-performing past work (top). In addition we conduct ablation studies (bottom) on novel frames with unseen poses. Our full method outperforms related work and ablated results of our hybrid representation.}
\label{tab:4_metrics}
\begin{tabular}{@{}l|cc@{}}  
\toprule
Method & FSIM$\uparrow$ & LPIPS ($\times10^{-2}$) $\downarrow$ \\ \midrule
SCARF~\cite{Feng2022scarf} & 0.764 & 5.00  \\
Animatable Gaussians~\cite{li2024animatablegaussians} & 0.827 & 3.39 \\ \midrule
3DGS-Only & 0.825 & 3.41 \\ 
PBR-Only & 0.809 & 4.67 \\
Ours & \textbf{0.834} & \textbf{3.38} \\ \bottomrule
\end{tabular}
\vspace{-0.1cm}
\end{table}

%% file: figtext/4_ablations.tex
\begin{figure}[t]
  \centering
  \includegraphics[width=0.98\linewidth]{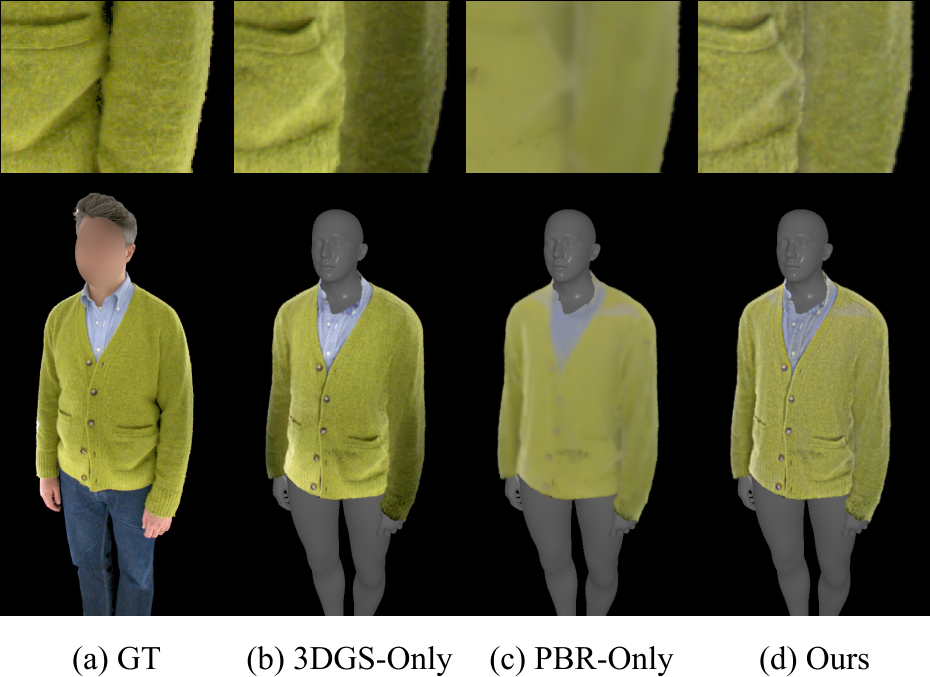}
   \caption{\textbf{Ablation}. We compare our approach to different ablations of our method and ground truth (a) of a novel pose. 
   3DGS-Only (b) has baked in shading from the training pose, resulting in unrealistic appearance when deformed to novel poses. PBR-Only (c) has reshading, but lacks detailed shading, and can only model flat 2D texture colors. Ours (d) has the best of both worlds.}
   \label{fig:4_ablations}
   \vspace{-0.15cm}
\end{figure}

%% file: figtext/4_decomposition.tex
\begin{figure}[t]
  \centering
  \includegraphics[width=0.98\linewidth]{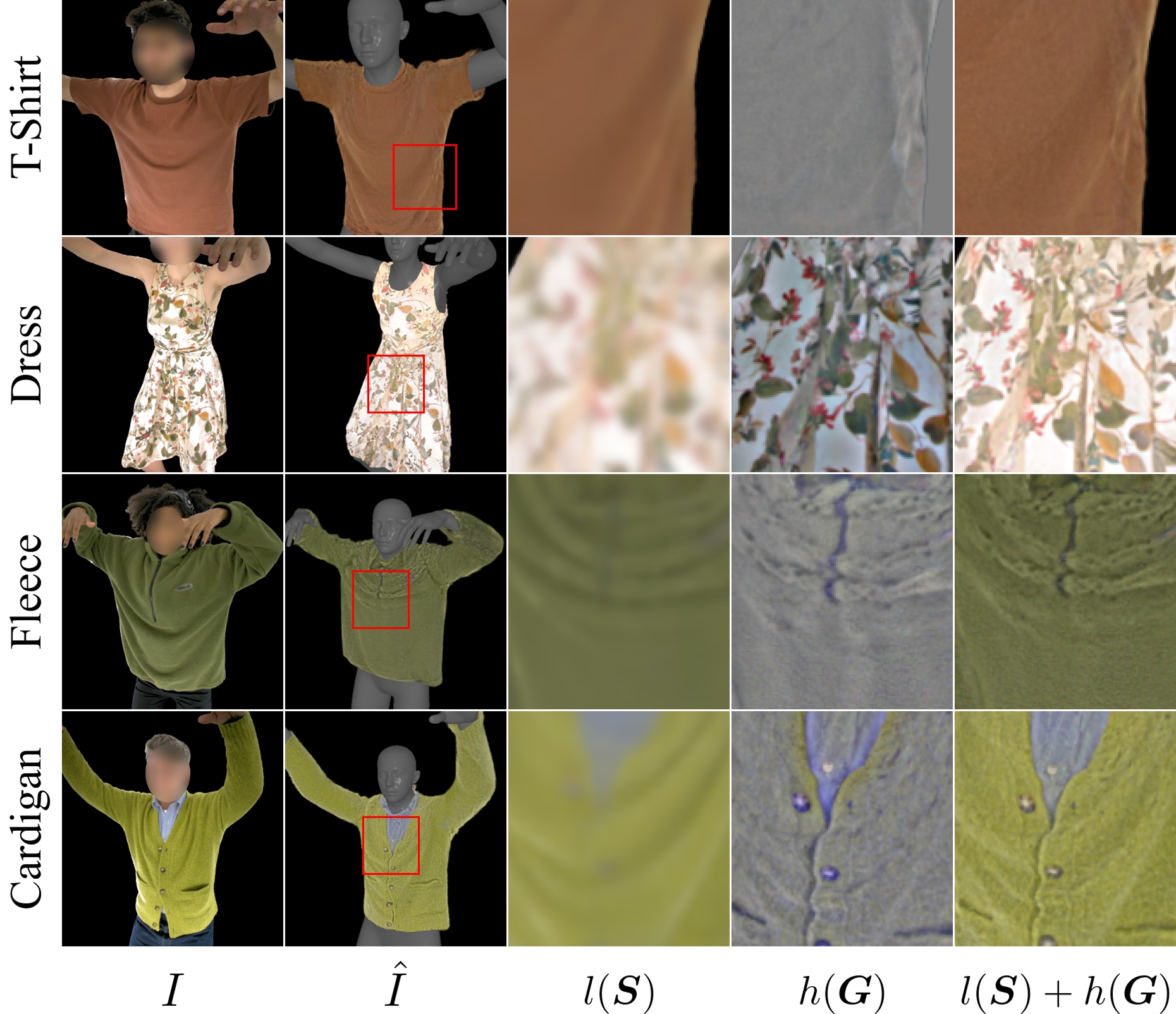}
  \caption{\textbf{Decomposition}. Our method composes the lowpass shaded Gaussians $\lowpass(\gshaded)$ with the highpass Gaussians $\highpass(\predimage)$ to render garments under novel deformations. Note that $\lowpass(\gshaded)$ contributes the far-field shading (\eg, mesh wrinkles) while $\highpass(\predimage)$ contributes the near-field, high-frequency details.}
   \label{fig:4_decomposition}
\end{figure}

%% file: figtext/4_baselines.tex
\begin{figure}[t]
  \centering
  \includegraphics[width=0.98\linewidth]{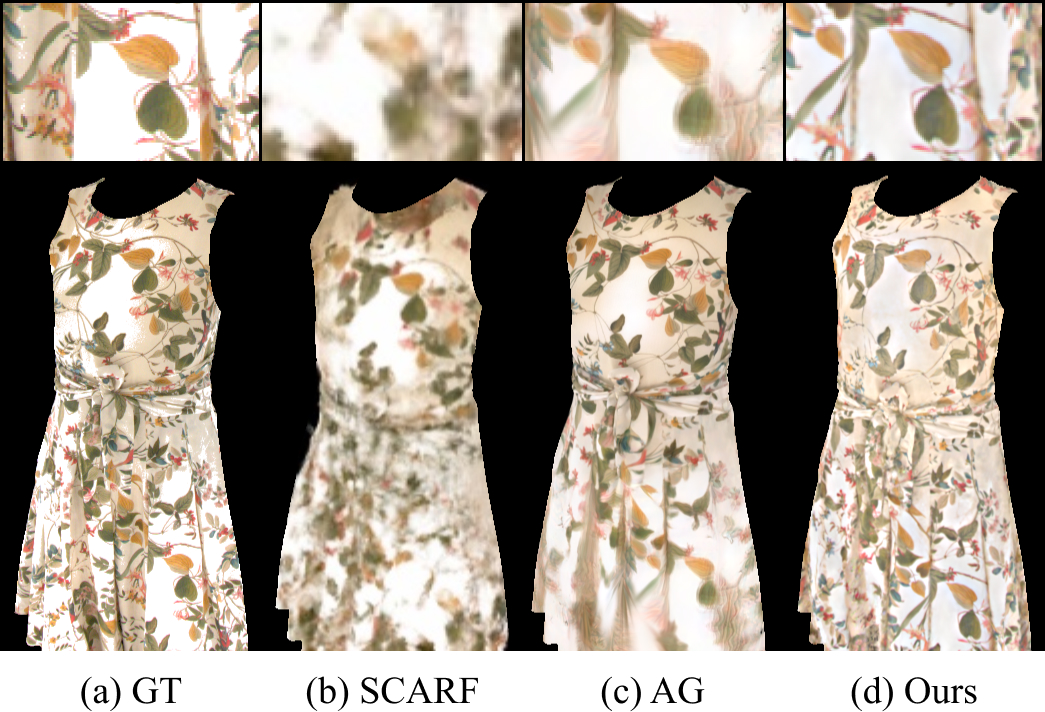}
   \caption{\textbf{Qualitative Comparisons}. We compare our proposed approach (d) with the state-of-the-art methods SCARF~\cite{Feng2022scarf} (b) Animatable Gaussians~\cite{li2024animatablegaussians} (c) in a novel pose setting. 
   We observe that both baselines struggle with loose clothing and at the same time they are not lighting-aware while our method can faithfully represent the dress and render it under novel lighting.}
   \label{fig:4_baselines}
   \vspace{-0.3cm}
\end{figure}

%% file: sec/5_discussion.tex
\vspace{0.15cm}
\subsection{Limitations and Future Work}\label{ssec:limitations}

\noindent\textbf{Pose-dependent Details and Relightability.} Our approach ignores the effects of deformation and novel lighting on high-frequency appearance. 
It is also limited by the capacity of 3DGS of generalizing to under-observed areas (\eg armpit area). While this can be partially inferred on a full garment by tracking multi-view video capture per garment as in~\cite{rong2024gaussiangarments}, another approach would be to capture fine details using close-up captures of fabrics and with varying lighting conditions for a fully relightable model.

\noindent\textbf{Albedo Extraction and Back-projection.} Our method uses an off-the-shelf albedo extraction technique~\cite{careagaIntrinsic}, which we apply per camera view, that generates inconsistent colors when back-projected onto the mesh. This causes some artifacts in the far-field render. Addressing this issue would enable us to make use of a smaller blur kernel and rely more heavily on the generalizable far-field PBR component. 
\noindent\textbf{Range of Fabrics.} We target common fabrics with fuzzy surfaces that are difficult to reproduce with pure PBR methods. Long fur or transparent fabrics would require mesh-reconstruction techniques and accurate tracking \cite{rong2024gaussiangarments}.

\input{figtext/4_relighting}

\noindent\textbf{Geometry Misalignment.} To better match ground truth geometry, we can leverage differentiable simulation~\cite{stuyck2023diffxpbd, li2023diffavatar} to recover more accurate rest shape and material parameters. This would enable editability,  resizing~\cite{chen2024dress}, and sag-free rest shape estimation~\cite{Hsu2022} to obtain simulated garments that remain consistent in size with the reconstructed geometries. 
This would improve feature alignment at test time, potentially improving similarity metrics to ground truth. 
Finally, our method assumes shell geometry along the entire garment, which would not accurately model the deformation of pockets, multi-layer garments, or garment openings.

\noindent\textbf{Runtime}: 
Once reconstructed, our garments can be leveraged in real-time applications such as games or virtual telepresence. Our method relies exclusively on techniques that run in real-time: a) PBR techniques are commonly used in real-time games~\cite{akenine2019real}, and b) numerous recent works~\cite{wu20244d, niemeyer2024radsplat, peng2024rtg} demonstrate real-time Gaussian splat rendering. 
This, in combination with real-time simulation~\cite{macklin2016xpbd}, enable our garments to be rendered and simulated in real-time, making them highly versatile for various applications.

%% file: figtext/4_relighting.tex
\begin{figure}[t]
  \centering
  \includegraphics[width=0.98\linewidth]{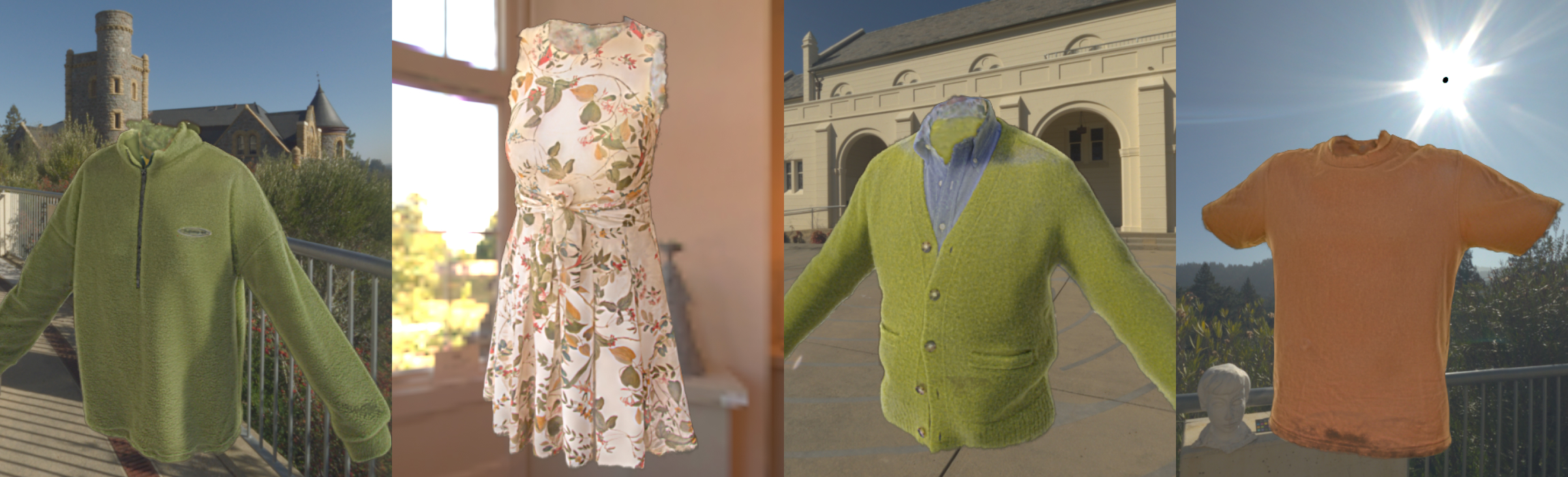}   \caption{\textbf{Relighting}. The PBR component of the composed render enables our assets to correctly respond to different lighting. 
  }
   \label{fig:4_relighting}
   \vspace{-0.3cm}
\end{figure}

%% file: sec/6_conclusion.tex
\vspace{0.23cm}
\section{Conclusion}
We introduced a new method that generates real-time capable virtual garment assets from a single frame multi-view capture. 
Our approach extends 3D Gaussian splatting applied to garment captures with the ability to generalize to new deformations such as those produced by cloth simulation tools and new lighting conditions. 
Our results outperform both qualitatively and quantitatively recent state-of-the-art methods while remaining simple and requiring only captures of a single pose. 
This eliminates the need for complex surface tracking and prevents blurring of fine details due to averaging and motion blur effects of multi-frame captures. 
We believe that this work brings us one step closer to more realistic clothed avatars with garment assets suitable for virtual try-on and telepresence applications.

%% file: sec/X_suppl.tex
\clearpage
\setcounter{page}{1}
\setcounter{section}{0}
\maketitlesupplementary

\section*{A. Qualitative Results and Video}
Please see the accompanying supplementary video for an introduction and walkthrough of our approach, as well as additional qualitative results.
This video shows renders of our simulation-ready garments on unseen motion sequences with outdoor and indoor environment maps.
This demonstrates the ability of our method to generalize to novel poses and various lighting conditions.
We provide a side-by-side comparison against the original ``3D Gaussian Splatting'' (3DGS-Only) method. Not only is the shading baked into the garments produced by 3DGS-Only, they are also not relightable. Lastly, we demonstrate 360 relighting, where the garments are dynamically relit with a rotating illumination map.

\section*{B. Mesh-Embedded 3DGS Reconstruction}

In this section, we provide additional details on the mesh-embedded 3DGS representation and optimization approach.

\myparagraph{Mesh Reconstruction}
We remesh the garment to 11-18k vertices using Houdini~\cite{Houdini}, though any remeshing software would suffice.
After remeshing, we unwrap the mesh in Blender~\cite{Blender} to obtain the UV map.

\myparagraph{Mesh-embedded Gaussian Splat Representation}
Once we have sampled Gaussians on the mesh surface, we follow Qian et al.~\cite{qian2024gaussianavatars} to define each Gaussian in the local coordinate frame of its parent triangle on the mesh. This allows us to deform the Gaussians as the mesh deforms at test time. The origin of the local coordinate frame $\tposition$ is the centroid of the triangle. The face orientation is defined with three vectors: one of the edges normalized, the normal vector of the triangle, and their cross product. They are concatenated as column vectors to form the rotation matrix $\trotation \in \text{SO}(3)$.

We represent each Gaussian's rotation quaternion $\grotation\in\Hamilton$, position $\gposition \in \R^3$, and scale $\gscale\in\R^3_+$ in the local coordinate frame of its parent triangle. To render the Gaussians, we transform the Gaussian parameters into world coordinates with the following equations:
\begin{align}
    \grotation' &= \bm{R}\grotation, \\
    \gposition' &= k\bm{R}\gposition + \bm{\tau}, \\
    \gscale' &= k\gscale,
\end{align}
where $k\in\R_+$ is the scale of the triangle defined as $\frac{B+H}{2}$, where $B$ and $H$ are the base and height of the triangle, respectively. 

\myparagraph{3DGS Initialization}
We initialize our Gaussian model with 1 million splats on the surface of the reconstructed full-body mesh. Note that only the subset of splats that belong to the garment segmented mesh are kept for inference, so in practice we use much fewer than 1 million splats to model the garment.
Following~\cite{kerbl2023Dgs} we initialize opacity $\gopacity$ to 0.1, and initialize only the first three coefficients of spherical harmonics $\gsphs$ using randomly sampled RGB color values, setting the remaining coefficients to zeros.
We sample RGB colors uniformly from $\grgb \sim \mathcal{U}(0.5, 0.7)^3$ where $\grgb \in [0,1]^3$ is the RGB color of the Gaussian splat.
The local Gaussian position is computed by transforming the sampled splat positions from world to local coordinates. Following~\cite{qian2024gaussianavatars}, we initialize the local Gaussians with identity rotation and unit scale $\gscale$.

\myparagraph{3DGS Optimization}
We apply a mask regularization loss term as described in
\supplRefMeshEmbeddedGaussians
. The kernel size $K_\text{fg}$ is computed as the image height scaled by a mask erosion factor $\gamma_\text{fg}$. The loss is weighted by $\lambda_{\text{fg}}$.

Our optimization largely follows the settings from~\cite{kerbl2023Dgs}, with some modifications to the values. The full list of optimization parameters and their values are provided in \cref{tab:X_3dgs_params}.
In addition to the original implementation, we also provide the hyperparameters to additional modules included in our implementation in \cref{tab:X_3dgs_params_ours}.

\input{table/X_3dgs_params}
\input{table/X_3dgs_params_ours}

\myparagraph{Caveats}
After 3DGS optimization described in
\supplRefMeshEmbeddedGaussians
, the mesh is culled to remove triangles deemed to not belong to the garment of interest. Due to the nature of global 3DGS optimization, some Gaussians that contributed to the color of the garment may belong to triangles representing other parts of the reconstructed body mesh. Thus culling triangles after 3DGS optimization subtly reduce the opacity in some areas of the garment (\eg bottom of the T-shirt as visible in \cref{fig:X_novel_motion_and_lighting} or beside the left pocket of the cardigan in \supplRefTeaser). We believe this is a minor issue and not a fundamental limitation of the method. For instance, to improve surface opaqueness for each pixel, we may eliminate contributions from Gaussians on back-facing triangles representing the interior of the garment, thus resulting in a single front facing cloth layer remaining, responsible for the foreground reconstruction during the 3DGS optimization step.

\section*{C. PBR Appearance Reconstruction}

We provide implementation details for the comparisons with Lambertian and Disney BRDF~\cite{burley2012physically} PBR models in
\supplRefPbr~(b) and (c), as well as our PBR model (d) from the main document.
We use Mitsuba~\cite{Mitsuba3} as our differentiable renderer.
We report hyperparameters for optimization in \cref{tab:X_mitsuba_params}.
We use a constant lighting model to approximate the overall lighting employed by our capture system, as described in 
\supplRefExperiments.
The base color generated by the neural network (\supplRefPbrAppearance), despite being free from baked-in lighting and shadowing, could still contain colors influenced by the lights.
To account for such color scaling, we additionally optimize the RGB radiance of the constant lighting model.
We report the initial and optimized parameters per model in \cref{tab:X_mitsuba_init} and \cref{tab:X_mitsuba_opt}, respectively. We also report PSNR values for the different PBR models, per garment, in \cref{tab:X_mitsuba_metrics}.

\input{table/X_mitsuba_params}
\input{table/X_mitsuba_init}
\input{table/X_mitsuba_opt}
\input{table/X_mitsuba_metrics}

\section*{D. Gaussian---PBR Hybrid Rendering}

\myparagraph{Reconstruction Error}
Dropping the subscript $t$ for simplicity, we can write the reconstruction error $\epsilon$ (when compared to the real world images) as
\begin{align}
    \epsilon &= \gtimage - \reconstruction \notag \\
    &= \gtimage - (\highpass(\predimage) + \lowpass(\gshaded)) \notag \\
    &= \gtimage - (\predimage - \lowpass(\predimage)) - \lowpass(\gshaded) \notag \\
    &= \underbrace{\lowpass(\predimage) - \lowpass(\gshaded)}_{\text{additional error}} + \underbrace{\gtimage - \predimage}_{\text{3DGS error}}. \label{eq:recerr}
\end{align}
For training frames, given that $\gtimage - \predimage$ is small, this suggests that the error mostly occurs in PBR reconstruction. It also suggests that it is sufficient for the PBR reconstruction to approximate the ground truth images only in the low frequency mode.
At test time, $\gtimage - \predimage$ is no longer small due to baked-in lighting and shadowing in $\predimage$. Intuitively, however, we expect novel shading missing from $\predimage$ to be reintroduced by $\lowpass(\gshaded) - \lowpass(\predimage)$. Additional error could be introduced at test time from mismatched geometry since our method does not rely on tracking.

\section*{E. Experiments}

\myparagraph{Optimization and Inference Efficiency}
We report performance and runtime efficiency in \cref{tab:X_performance}.
All 3DGS rendering is performed on an NVIDIA A100 GPU, while all PBR rendering and simulation is run on an NVIDIA RTX 3080 GPU. Note that while the reported timing is not real time, each of the components have real-time counterparts, making the method compatible with real-time pipelines.

\input{table/X_performance}

\myparagraph{Simulation}
We use eXtended Position Based Dynamics~(XPBD)~\cite{macklin2016xpbd} as our choice of simulator.
The simulation parameters are provided in \cref{tab:X_sim_params}. In all examples presented in the main paper, the material parameters were kept constant. However, to better approximate the behavior of thicker garments like the fleece and cardigan we set the bending stiffness to be x10 larger than what is used for the t-shirt and dress simulations in \cref{fig:X_novel_motion_and_lighting}.

\input{table/X_sim_params}
\myparagraph{Baseline Setup} For Animatable Gaussians, we train on all views with a video sequence of diverse poses containing roughly 8k frames. 
SCARF is a NeRF-based method that reconstructs animatable clothed humans from monocular video.
We train on the same training data as was used for our method, excluding extreme camera poses (because PIXIE, the method for SMPL body estimation, struggles with extreme camera poses 
\cite{yao2021pixie}
). Although it is a method for monocular video, we reframe our multi-view static pose setting as a monocular video by concatenating all the views into a video, as done in~\cite{rong2024gaussiangarments}.
Animatable Gaussians takes 3 days to train. SCARF takes 14 hours.

\myparagraph{Additional Metrics} We report additional metrics, structural similarity (SSIM)~\cite{zhou2004image} and peak signal-to-noise ratio (PSNR), comparing with baselines and ablations in \cref{tab:X_metrics}. Note that our method does not optimize for the garment rest shape. Instead, we use the reconstructed geometry as the rest shape for simulation. While our results are more crisp and produce more realistic dynamics (see \supplRefBaselines), our method can produce sagging and in general is not guaranteed to match the geometry from ground truth frames in novel poses. We believe this is the reason why our method (and ablated versions of it) achieve lower scores on metrics that are sensitive to pixel alignment. Nevertheless, we report these metrics for reference.

\input{table/X_metrics}
\input{figtext/X_blur}
\input{figtext/X_actorshq}

\myparagraph{Novel motion} We evaluate the generalization of our garments to novel motions in \cref{fig:X_novel_motion_and_lighting}. Please see the supplementary video to view the motion sequences.
\input{figtext/X_novel_motion_and_lighting}

\myparagraph{Lighting 360} We show our garments relit under different rotations of the environment in \cref{fig:X_lighting_360}. Please see the supplementary video for additional renders.

\myparagraph{Blur Analysis} We analyze the effect of blur kernel size on training reconstruction for the \textit{cardigan} example. 
We report feature similarity (FSIM)~\cite{zhang2011fsim} index of our reconstruction in relation to the blur kernel size in \cref{fig:X_blur}.
Large amount of blur reduces the reconstruction error (\cref{eq:recerr}) since it is less challenging for PBR reconstruction to approximate an overly blurred image. However, in the meantime, this introduces more contribution from 3DGS, which prevents us from generalizing to novel poses at testing time. 
To strike a balance, we select a blur kernel size $71\times71$ for our training image with resolution $4096\times2668$. We use the same kernel size for all our examples, however the optimal blur kernel size ought to vary by garment type.
Garments with high-frequency details (thick wool knits) necessitate larger 
kernels to accurately capture these intricate features. One way to automate the process of choosing a garment-dependent 
kernel size is by progressively blurring the input image until 
its gradient magnitudes fall below a threshold, indicating 
that high-frequency details have been removed. We leave this extension to future work.

\myparagraph{ActorsHQ dataset} We show that our method works on the ActorsHQ dataset in \cref{fig:X_actorshq} showing the reconstruction for an unseen pose. The quality metrics FSIM=0.938, LPIPS=0.0227, PSNR=33.99, SSIM=0.977 are similar to those in the main paper and
slightly outperform PhysAvatar for the selected frame.

\input{figtext/X_lighting_360}

%% file: table/X_3dgs_params.tex
\begin{table}[h]
\centering
\caption{\textbf{3DGS Optimization Settings}: We provide the optimization parameters and their values for 3DGS optimization. The parameters are derived from the original 3DGS implementation~\cite{kerbl2023Dgs}, with some modifications to the values.}
\label{tab:X_3dgs_params}
\begin{tabularx}{\columnwidth}{l@{\hskip 0.5in}X}
\toprule
Parameter & Setting \\ \midrule
num. optimization iterations & 30000 \\
optimizer & Adam \\
position learning rate (init) & 0.5e-4 \\
position learning rate (final) & 1.0e-7 \\
position learning rate (max steps) & 30000 \\
feature learning rate & 0.005 \\
opacity learning rate & 0.005 \\
scaling learning rate & 0.005 \\
rotation learning rate & 0.005 \\
SSIM loss weight $\lambda_\text{SSIM}$ & 0.2 \\
SH increase frequency & 500 \\ \bottomrule
\end{tabularx}
\end{table}

%% file: table/X_3dgs_params_ours.tex
\begin{table}[ht]
    \centering
    \caption{\textbf{Additional 3DGS Optimization Settings}: We provide the optimization parameters and settings for additional modules introduced in our implementation below.}
    \label{tab:X_3dgs_params_ours}
    \begin{tabularx}{\columnwidth}{l@{\hskip 0.5in}X}
        \toprule
        Parameter                                & Setting \\ \midrule
        mask erosion factor $\gamma_{\text{fg}}$ & 0.03    \\
        mask loss weight $\lambda_{\text{fg}}$   & 0.1     \\
        \bottomrule
    \end{tabularx}
\end{table}

%% file: table/X_mitsuba_params.tex
\begin{table}[h]
\centering
\caption{\textbf{Optimization Settings for Shading Reconstruction}: We provide the optimization parameters for shading reconstruction, which are implemented in the Mitsuba~\cite{Mitsuba3} differentiable renderer.}
\label{tab:X_mitsuba_params}
\begin{tabularx}{\columnwidth}{l@{\hskip 0.5in}X}
\toprule
Parameter & Setting \\ \midrule
num. optimization iterations & 3000 \\
optimizer & Adam \\
learning rate & 0.01 \\
samples per pixel & 4 \\
integrator type & path \\ 
lighting model & constant \\ \bottomrule
\end{tabularx}
\end{table}

%% file: table/X_mitsuba_init.tex
\begin{table}[ht]
    \small
    \centering
    \caption{\textbf{PBR Parameter Initialization}: We report the initializations used for each PBR model. Parameters that are not listed use the default values in Mitsuba.}
    \label{tab:X_mitsuba_init}
    \begin{tabularx}{\columnwidth}{lXc}
        \toprule
        PBR Model & Parameter         & Setting            \\
        \midrule
        \multirow{1}{*}{Lambertian}
                  & lighting radiance & (1.00, 1.00, 1.00) \\
        \midrule
        \multirow{4}{*}{Disney BRDF~\cite{burley2012physically}}
                  & lighting radiance & (1.00, 1.00, 1.00) \\
                  & roughness         & 0.8                \\
                  & sheen             & 1.4                \\
                  & specular          & 0.0                \\
        \midrule
        \multirow{6}{*}{Ours}
                  & lighting radiance & (1.00, 1.00, 1.00) \\
                  & roughness         & 0.8                \\
                  & sheen             & 1.4                \\
                  & sheen color       & (0.75, 0.73, 0.27) \\
                  & sheen roughness   & 0.5                \\
                  & specular          & 0.0                \\
        \bottomrule
    \end{tabularx}
\end{table}

%% file: table/X_mitsuba_opt.tex
\begin{table*}[ht]
    \centering
    \caption{\textbf{Optimized PBR Parameters}: We report the optimized parameters per PBR-model, for each garment.}
    \label{tab:X_mitsuba_opt}
    \begin{tabularx}{\textwidth}{lXcccc}
        \toprule
                  &                   & \multicolumn{4}{c}{Garments}                                                                \\
        \cmidrule(lr){3-6}
        PBR Model & Parameter         & T-Shirt                      & Dress              & Fleece             & Cardigan           \\
        \midrule
        \multirow{1}{*}{Lambertian}
                  & lighting radiance & (4.15, 5.16, 5.27)           & (1.39, 1.43, 1.44) & (7.07, 6.82, 8.18) & (3.56, 3.59, 5.53) \\
        \midrule
        \multirow{4}{*}{Disney BRDF~\cite{burley2012physically}}
                  & lighting radiance & (3.82, 7.27, 7.56)           & (1.32, 1.44, 1.51) & (6.32, 5.97, 6.23) & (3.00, 3.02, 4.02) \\
                  & roughness         & 0.09                         & 0.00               & 1.00               & 0.85               \\
                  & sheen             & 0.56                         & 4.50               & 0.57               & 1.46               \\
                  & sheen tint        & 0.66                         & 0.00               & 0.39               & 0.56               \\
        \midrule
        \multirow{6}{*}{Ours}
                  & lighting radiance & (2.85, 3.34, 3.31)           & (1.02, 1.12, 1.24) & (4.86, 4.75, 5.01) & (2.88, 2.90, 4.62) \\
                  & roughness         & 0.97                         & 0.53               & 1.00               & 0.99               \\
                  & sheen             & 0.35                         & 1.05               & 0.22               & 0.40               \\
                  & sheen color       & (0.35, 0.22, 0.15)           & (0.51, 0.48, 0.34) & (0.23, 0.24, 0.13) & (0.39, 0.38, 0.13) \\
                  & sheen roughness   & 0.66                         & 0.76               & 0.62               & 0.42               \\
        \bottomrule
    \end{tabularx}
\end{table*}

%% file: table/X_mitsuba_metrics.tex
\begin{table}[ht]
    \centering
    \caption{\textbf{Quantitative Results on Shading Reconstruction}: We report PSNR per-PBR model, on each garment.}
    \label{tab:X_mitsuba_metrics}
    \begin{tabular}{@{}l|cccc@{}}
        \toprule
        PBR Model                               & T-Shirt        & Dress          & Fleece         & Cardigan       \\ \midrule
        Lambertian                              & 34.76          & 21.94          & 33.01          & 26.55          \\
        Disney BRDF~\cite{burley2012physically} & 37.57          & 22.26          & 34.29          & 27.10          \\
        Ours                                    & \textbf{37.67} & \textbf{22.93} & \textbf{34.40} & \textbf{27.23} \\ \bottomrule
    \end{tabular}
\end{table}

%% file: table/X_performance.tex
\begin{table}[ht]
    \centering
    \caption{\textbf{Optimization and Inference Efficiency}: We report the efficiency for both optimization and inference of our method.}
    \label{tab:X_performance}
    \begin{tabularx}{\columnwidth}{lXc}
        \toprule
        Stage & Module            & Runtime  \\
        \midrule
        \multirow{2}{*}{Optimization}
              & 3DGS optimization & 3h       \\
              & PBR optimization  & 12m      \\
        \midrule
        \multirow{4}{*}{Inference}
              & XPBD Simulation   & 11 FPS   \\
              & PBR rendering     & 3.2 FPS  \\
              & 3DGS rendering    & 1.6 FPS  \\
              & Image filtering   & 14.4 FPS \\
        \bottomrule
    \end{tabularx}
\end{table}

%% file: table/X_sim_params.tex
\begin{table}[ht]
    \centering
    \caption{\textbf{Simulation Parameters}: We provide the parameters that we use for garment simulation. We use the XPBD simulator~\cite{macklin2016xpbd} with spring constraints for both stretching and bending.}
    \label{tab:X_sim_params}
    \begin{tabularx}{\columnwidth}{l@{\hskip 0.5in}X}
        \toprule
        Parameter          & Setting \\ \midrule
        cloth-body offset  & 0.4 cm  \\
        frame rate         & 30      \\
        substeps per frame & 30      \\
        XPBD iterations    & 20      \\
        \bottomrule
    \end{tabularx}
\end{table}

%% file: table/X_metrics.tex
\begin{table}[h]
\centering
\caption{\textbf{Quantitative Comparisons}: We report additional metrics comparing with existing work and ablated versions of our method. Note that our method (and ablations) do not optimize for the garment rest shape. We believe this is the reason that baselines outperform on metrics that are sensitive to pixel alignment.}
\label{tab:X_metrics}
\begin{tabular}{@{}l|cc@{}}
\toprule
Method & SSIM$\uparrow$ & PSNR$\uparrow$ \\ \midrule
SCARF~\cite{Feng2022scarf} & 0.937 & 25.96  \\
Animatable Gaussians~\cite{li2024animatablegaussians} & 0.945 & \textbf{29.75} \\ \midrule
3DGS-Only & 0.938 & 28.62 \\
PBR-Only & \textbf{0.947} & 28.65 \\
Ours & 0.939 & 28.37 \\ \bottomrule
\end{tabular}
\end{table}

%% file: figtext/X_blur.tex
\begin{figure}[t]
  \centering
  \includegraphics[width=0.98\linewidth]{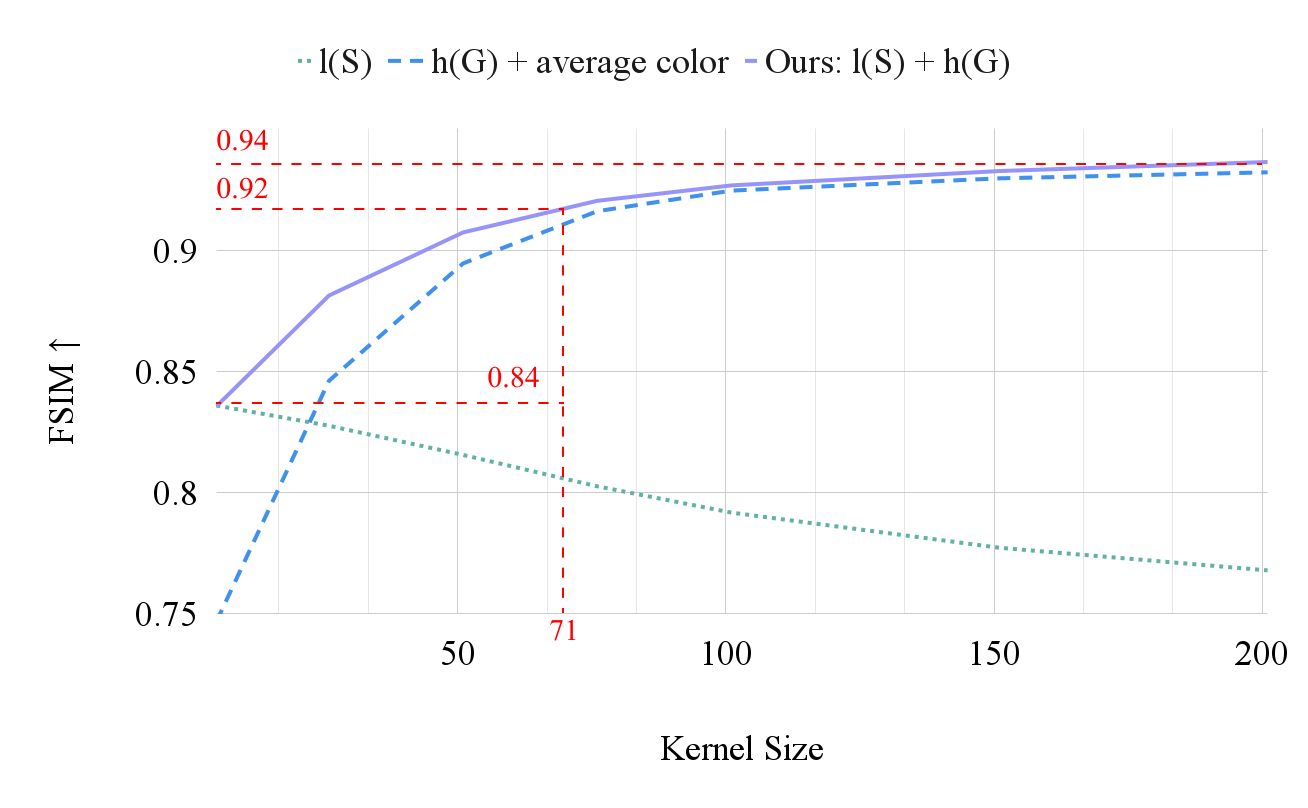}   \caption{\textbf{Blur Analysis}. We analyze the reconstruction performance on the training frame of the \textit{cardigan} garment, across different levels of blur. Compared to no blur, our selected kernel size of 71 pixels (shown in red) improves the FSIM metric by 8\%. Increasing the blur kernel further produces no more than a 2\% increase in FSIM, while decreasing the generalizability of the appearance model to novel poses. FSIM for $l(\gshaded)$ and $h(\predimage) + $average color are plotted for reference showing that our model produces quantitatively better results at all blur levels.
  }
   \label{fig:X_blur}
\end{figure}

%% file: figtext/X_actorshq.tex
\begin{figure}[t]
  \centering
  \includegraphics[width=0.98\linewidth]{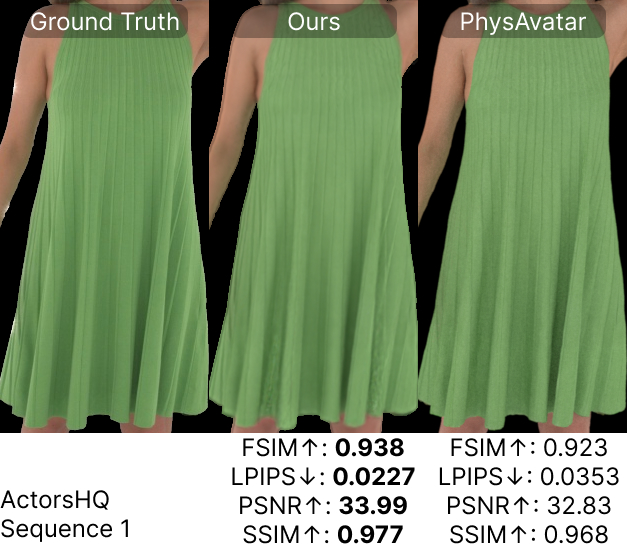}   \caption{\textbf{ActorsHQ example}. This figure demonstrates our method applied to the popular ActorsHQ dataset.
  }
   \label{fig:X_actorshq}
\end{figure}

%% file: figtext/X_novel_motion_and_lighting.tex
\begin{figure*}[t]
  \centering
  \includegraphics[width=0.98\linewidth]{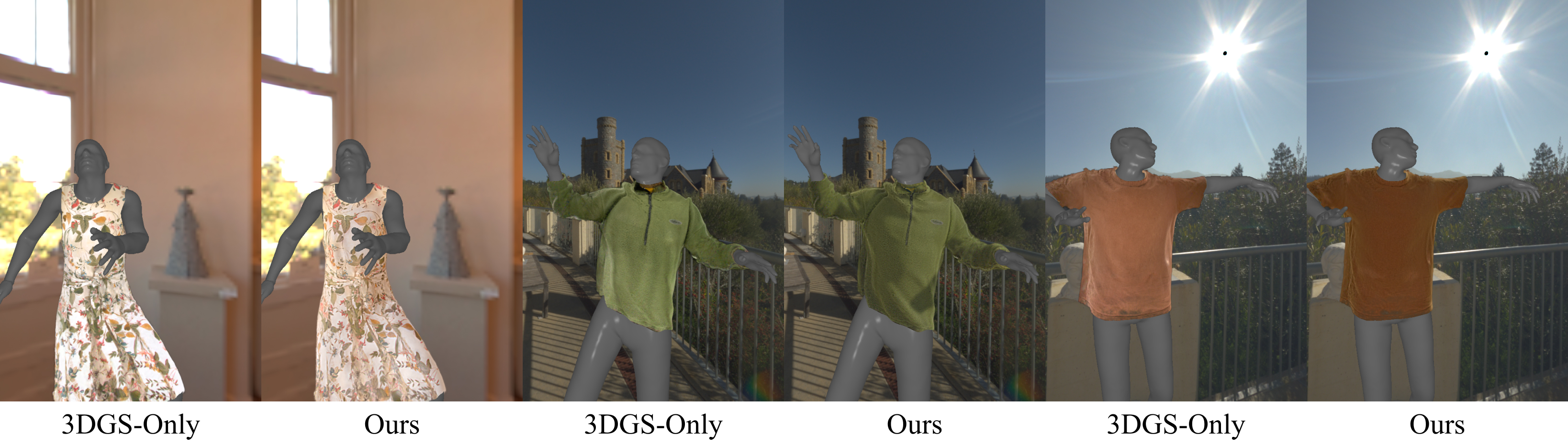}   \caption{\textbf{Novel Motion}. We evaluate our method on novel motion sequences.
  }
   \label{fig:X_novel_motion_and_lighting}
\end{figure*}

%% file: figtext/X_lighting_360.tex
\begin{figure*}[t]
  \centering
  \includegraphics[width=0.98\linewidth]{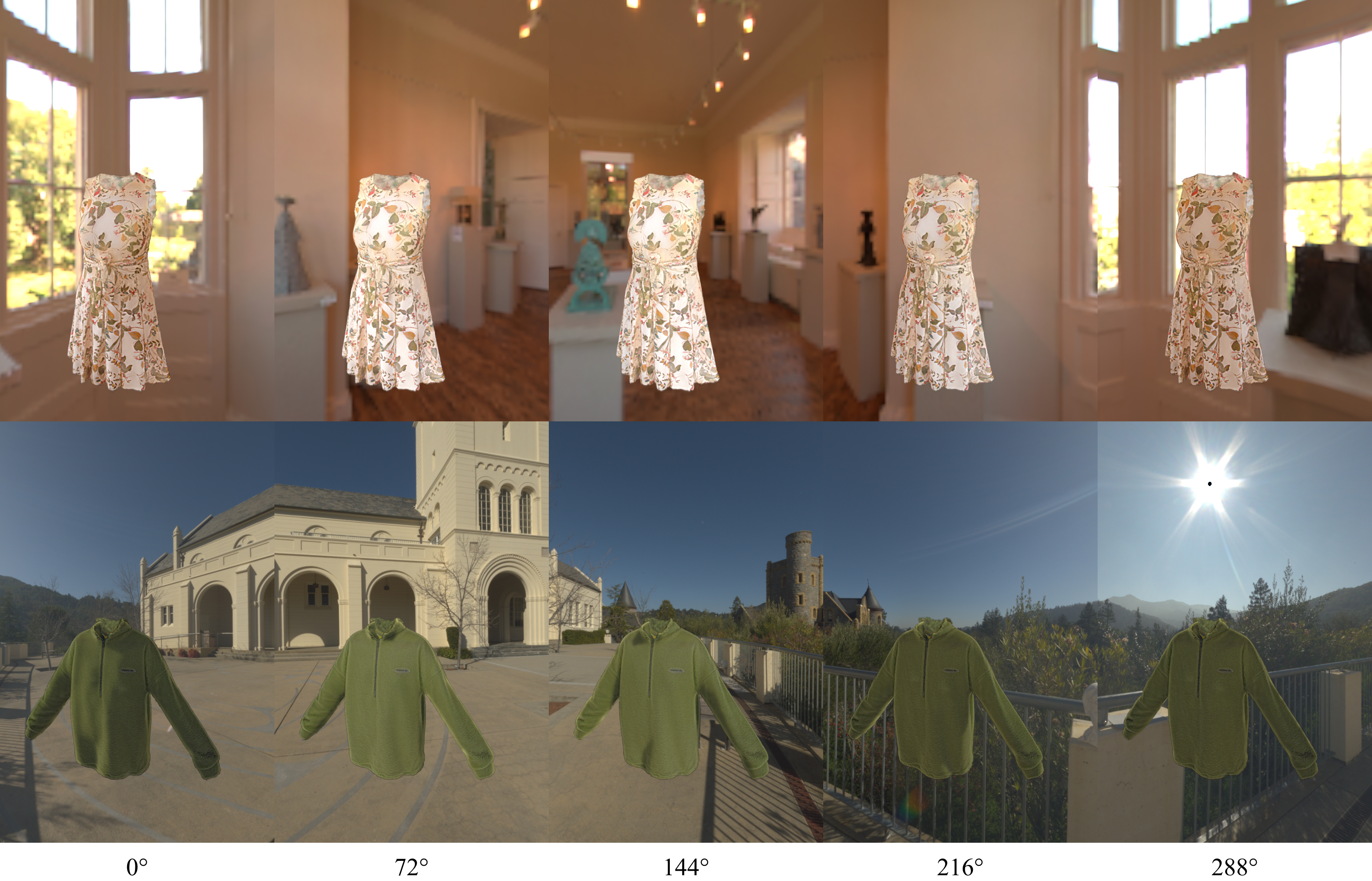}   \caption{\textbf{Lighting 360$^\circ$}. We show our garments under different rotations of the environment map.
  }
   \label{fig:X_lighting_360}
\end{figure*}